\newif\iflong
\longtrue

\iflong
	\newcommand{\shortorlong}[2]{#2}
\else
	\newcommand{\shortorlong}[2]{#1}
\fi

\shortorlong{%
    \documentclass[sigconf]{acmart}%
}{%
    \documentclass[sigconf,nonacm]{acmart}%
}

\usepackage[dvipsnames]{xcolor}
\usepackage{tikz}
\usepackage{tikzscale}

\usepackage{tikzuml}
\usepackage{tikztgg}

\usetikzlibrary{positioning, calc}

\usepackage[nolist]{acronym}

\usepackage{mathtools}

\usepackage{tabularx}

\usepackage[inline,shortlabels]{enumitem}

\usepackage{csquotes}

\usepackage[capitalise]{cleveref}
\crefname{equation}{Query}{Queries}
\Crefname{equation}{Query}{Queries}

\AtBeginDocument{%
  }

\copyrightyear{2026}
\acmYear{2026}
\setcopyright{cc}
\setcctype{by}
\shortorlong{%
    \acmConference[MODELS Companion 2026]{ACM/IEEE 29th International Conference on Model Driven Engineering Languages and Systems}{October 04--09, 2026}{Málaga, Spain}
    \acmBooktitle{ACM/IEEE 29th International Conference on Model Driven Engineering Languages and Systems (MODELS Companion 2026), October 04--09, 2026, Málaga, Spain}
    \acmDOI{10.1145/3837062.3839053}
    \acmISBN{979-8-4007-2903-4/2026/10}%
}{}

\begin{acronym}
    \acro{tgg}[TGG]{triple graph grammar}
    \acro{lhs}[LHS]{left-hand side}
    \acro{rhs}[RHS]{right-hand side}
    \acro{nac}[NAC]{negative application condition}
\end{acronym}

\newcommand{\SrcGraph}[1]{#1_{\mathrm{S}}}
\newcommand{\TarGraph}[1]{#1_{\mathrm{T}}}
\newcommand{\CorrGraph}[1]{#1_{\mathrm{C}}}
\newcommand{\TrGraph}[1]{(#1,\allowbreak \SrcGraph{#1},\allowbreak \CorrGraph{#1},\allowbreak \TarGraph{#1})}
\newcommand{\RuleMorphism}{r\colon L \hookrightarrow R}
\newcommand{\RuleFull}{\rho = (\RuleMorphism, \mathit{NAC}, \mathcal{C})}
\newcommand{\TrafoStep}[4]{#1 \Longrightarrow_{#3, #4} #2}
\newcommand{\TrafoSeq}[2]{#1 \Longrightarrow^* #2}
\newcommand{\TrafoSeqRuleSet}[3]{#1 \Longrightarrow^*_{#3} #2}
\newcommand{\LangTGG}[1]{\mathcal{L}(#1)}

\newcommand{\Lind}{L_{\mathrm{ind}}}
\newcommand{\mind}{m_{\mathrm{ind}}}

\newcommand{\viewAttr}{\mathit{va}}
\newcommand{\helperAttr}{\mathit{ha}}
\newcommand{\readAttr}{\mathit{ra}}
\newcommand{\modelElement}[1]{\emph{#1}}

\DeclareMathOperator{\Median}{Median}
\DeclareMathOperator{\Mode}{Mode}

\newcommand{\matchOrTranslate}{%
    \raisebox{-4pt}{\tikz{
        \node[checked box] (block-out-source-match-0) {};
        \node[box] (block-out-source-match-1) [right=.7em of block-out-source-match-0] {};
        \node[checked box] (block-out-source-match-2) [right=1em of block-out-source-match-1] {};
        \node at ($(block-out-source-match-0)!.5!(block-out-source-match-1)$) {/};
        \draw[->] (block-out-source-match-1) -- (block-out-source-match-2);}%
    }
}

\begin{document}

\shortorlong{%
    \title[Extending Triple Graph Grammars for Model View Definitions]{Extending Triple Graph Grammars to Formalize Complex View Definitions on Families of Models}%
}{%
    \title[Extending Triple Graph Grammars for Model View Definitions]{Extending Triple Graph Grammars to Formalize Complex View Definitions on Families of Models -- Long Version}%
}

\author{Lars König}
\orcid{0000-0002-1751-1291}
\email{lars.koenig@kit.edu}
\affiliation{%
  \institution{Karlsruhe Institute of Technology (KIT)}
  \city{Karlsruhe}
  \country{Germany}
}

\author{Jens Kosiol}
\orcid{0000-0003-4733-2777}
\email{jens.kosiol@b-tu.de}
\affiliation{%
  \institution{Brandenburgische Technische Universität Cottbus -- Senftenberg}
  \city{Cottbus}
  \country{Germany}
}

\renewcommand{\shortauthors}{Lars König and Jens Kosiol}

\begin{abstract}
  View-based development is an important technique to manage complexity and facilitate cooperation in the development of modern and complex systems of systems. 
  This approach places high demands on view definition languages used in the development process. 
  First, such a language needs to be approachable and extremely versatile so that developers from diverse backgrounds are able to use it and can define views for a diverse set of different tasks. 
  Yet, the language also needs to have a precise and formal semantics so that it is possible to integrate the work performed on various views in a controlled manner to obtain a coherent overall system. 
  In this paper, we continue previous work of equipping the NeoJoin view definition language with a formal semantics that is based on \acp{tgg}. 
  This paves the way for obtaining automated and incremental synchronization procedures between models and views that come with high formal guarantees for their behavior. 
  Simultaneously, our work serves as a further case study of the expressivity and usability of \acp{tgg}. 
  We identify one gap, namely convenient support for the translation of overlapping queries from the view definition language, and tackle that gap by introducing a skip semantics for \ac{tgg} rules, i.e., allowing to skip certain actions a rule prescribes, depending on the application context.
\end{abstract}




\maketitle
\acresetall

\section{Introduction}
\label{sec:intro}

Views are an important part of model-driven development to manage the complexity and consistency of models.
Both are frequent issues in the development of cyber-physical systems~\cite{feichtinger_industry_2022}.
View definition languages are used to define the available view types, i.e., view metamodels~\cite{goldschmidt_towards_2012}, and model-view transformations.
A number of view definition languages have been proposed~\cite{bruneliere_feature-based_2019}, however, these languages lack features desirable for the development of cyber-physical systems, such as expressive transformation operators, and bidirectional and incremental transformations~\cite{konig_towards_2025}.

The NeoJoin language~\cite{konig_language_2026} combines the definition of view types and model-view transformations in a query-style, SQL-like language.
Its syntax and transformation operators make it accessible for modeling experts as well as software and system engineers.
In order to provide a bidirectional and incremental implementation of the transformation operators, \citeauthor{konig_transforming_2025} proposed to generate \acp{tgg} from the view definitions~\cite{konig_transforming_2025}.
\Acp{tgg} provide a thorough formalism, including forward and backward operationalizations for the grammar's rules that can serve as the basis for automated and incremental synchronization processes.
Based on their work, we present a worked out application of \acp{tgg} for deriving model views, highlighting prospects and challenges of their application.
In detail, we extend the operators defined by \citeauthor{konig_language_2026} with three special cases: aggregation, cross-product joins, and overlapping queries.
We find that \acp{tgg} offer high formal guarantees but only limited support for the treatment of attributes and for parsing of the same model element repeatedly to represent it in different ways in the view.

As our second contribution, we extend \ac{tgg} rules by a \emph{skip semantics}, based on our idea of \emph{effect-oriented graph transformation}~\cite{kosiol_finding_2023}, and newly develop operationalized variants of \ac{tgg} rules with skip semantics.
We then use the defined skip semantics to suggest \ac{tgg} rules for overlapping queries. 
While this paper is confined to continue the foundational translation of queries from the NeoJoin language into \ac{tgg} rules, we at least briefly highlight which formal results we are confident to obtain based on this in the future.

We start by discussing related work in \cref{sec:related-work} and by introducing our running example and relevant background in \cref{sec:example,sec:background}. 
In \cref{sec:problems}, we present more complex queries from the NeoJoin language whose translation into \ac{tgg} rules remained open in~\cite{konig_towards_2025}, and we develop our solutions to these challenges in \cref{sec:solutions}.
In \cref{sec:conclusion}, we conclude, providing a prospect of the formal results that are now in reach, given the translations we develop in this paper.
In \shortorlong{%
    a long version of this paper~\cite{KK26},
}{%
    In \cref{app:examples-tggs,app:rule-operationalizations,app:aggregation}, 
}
we provide further technical details and additional examples we omit from the main paper for space reasons. 

\section{Related Work}
\label{sec:related-work}

Closest related to this paper are (i) other works developing view definition languages and (ii) case studies of \acp{tgg} in other contexts. 
Regarding (i), \citeauthor{bruneliere_feature-based_2019} provide a survey of model view approaches, however, only a limited number of them provides query-style view definitions with bidirectional and incremental model-view transformations~\cite{bruneliere_feature-based_2019}.
Most notable are OpenFlexo~\cite{golra_addressing_2016} and VIATRA Viewers~\cite{debreceni_query-driven_2014}.
In contrast to NeoJoin~\cite{konig_language_2026}, neither offers a unified DSL for the definition of view types and model-view transformations, with OpenFlexo providing multiple DSLs and VIATRA Viewers using annotations to define the view type.
Closely related to this work is the transformation approach taken by VIATRA Viewers, which builds on EMF-IncQuery~\cite{ujhelyi_emf-incquery_2015} and thus uses graph patterns as model queries.
The graph patterns are specified in a textual syntax and matched against a model instance using Rete networks~\cite{forgy_rete_1982}, which store partial matches and use these to recompute matches upon changes to the model. 
\Acp{tgg} have also already been employed to derive views of models~\cite{JakobKS06,AnjorinRDS14}, however, in these works of Jakob, Anjorin et al., the focus is on restricting the kinds of allowed \ac{tgg} rules (to gain efficiency) and to not materialize the view (to avoid data duplication) -- two topics not further considered in this paper. 

Regarding (ii), \acp{tgg} have been applied in a range of different scenarios; we here discuss papers that reflect on the pros and cons they experienced when using \acp{tgg}. 
\citeauthor{HermannGNEBMPEE14}~\cite{HermannGNEBMPEE14} used \acp{tgg} to translate satellite control procedures, \citeauthor{BlouinPDSD14}~\cite{BlouinPDSD14} to synchronize between a textual and a visual representation of models in the Architecture Analysis and Design Language; and \citeauthor{BuchmannW16}~\cite{BuchmannW16} to bidirectionally and incrementally synchronize between class diagrams and Java code. 
Generally, these applications of \acp{tgg} were successful and their formal underpinning and declarative and approachable nature proved to be advantages. 
Still, in each of the works the authors had to somehow modify or extend \acp{tgg}, in particular to tackle rule management and scalability issues.
Finally, \citeauthor{AnjorinB26}~\cite{AnjorinB26} investigated situations that are especially challenging for \acp{tgg}. 
They identified three situations, namely (i) ignoring certain elements; (ii) the treatment of attributes, in particular, when the same information is represented structurally in one model but numerically in the other; and (iii) related models that are to be parsed in opposite orders. 
With our introduction of a skip semantics for \ac{tgg} rules in \cref{sec:solutions:overlap} we provide a solution for their first challenging situation.

\section{Running Example}
\label{sec:example}

To illustrate the problems we are facing when implementing our query language with \acp{tgg}, we present a small example from the domain of system modeling and simulation.
In our example, we have two models: a blocks-and-ports model for simulating hardware components and a model describing the sensors available in our hardware components. 
\begin{figure}
    \centering
    \includegraphics{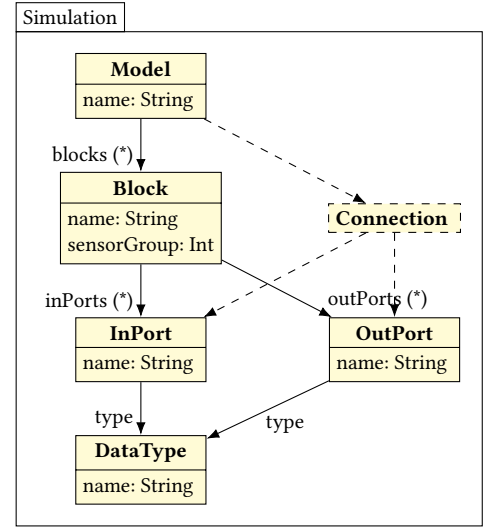}
    \caption{Example metamodel for the simulation of hardware components, consisting of connected blocks and ports. The references from and to the metaclass \emph{Connection} are dashed, as they are not relevant for our example.}
    \label{fig:example-simulink}
\end{figure}
We show the metamodel of our blocks-and-ports model in \cref{fig:example-simulink}.
The top level entity of this \emph{Simulation} metamodel is a \emph{Model}, which contains \emph{Blocks} and \emph{Connections}.
As the connections between blocks are not relevant for our examples, we will not explain them in more detail.
A \emph{Block}, on the other hand, has a \emph{name} and a number of input and output ports, represented by the classes \emph{InPort} and \emph{OutPort}.
Further, a \emph{Block} can be associated to multiple sensors, identified by the attribute \emph{sensorGroup}.
Multiple \emph{Blocks} can be associated to the same sensors, i.e., have the same value for the attribute \emph{sensorGroup}.
\emph{In-} and \emph{OutPorts} have a \emph{name} and a \emph{type}.
\begin{figure}
    \centering
    \includegraphics{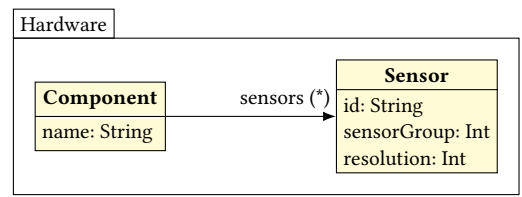}
    \caption{Example metamodel for sensor hardware.}
    \label{fig:example-hardware}
\end{figure}
In addition, we show our \emph{Hardware} metamodel, describing the available sensors, in \cref{fig:example-hardware}. 
A \emph{Component}, which represents the same thing as the \emph{Model} from our blocks-and-ports model, contains a number of \emph{Sensors}, which have an \emph{ID}, a \emph{resolution}, and are part of a \emph{sensorGroup}.

Notably, the system models for which we want to define and derive views consist of pairs, namely a blocks-and-ports and a sensor hardware model. 
This is, views project and combine their information to fulfill specific information needs. 
While, principally, such pairs (or, more generally, tuples) of models can be considered as a single model over a combined metamodel, the queries we address in this paper at least partially owe their complexity to the fact that they combine information from families of models into a single view.
As an example, consider the system models and view presented in \autoref{fig:view-join}.
In the view, we see all blocks associated to sensors with a resolution greater than 16.
Developers could use this view to analyze the blocks and check whether they support the high resolution data from the sensors.

\section{Background}
\label{sec:background}

In this section, we recall \acp{tgg} (\cref{sec:background:tgg}) and summarize which kind of model queries have already been translated into \ac{tgg} rules by König et al. in~\cite{konig_transforming_2025} (\cref{sec:background:operators}).

\subsection{Triple Graph Grammars}
\label{sec:background:tgg}

\Acfp{tgg} are an established formalism to specify the correspondence between models that are situated at different levels of abstraction~\cite{Schurr94, AnjorinLS15} and have already previously been used to specify views of models~\cite{JakobKS06, AnjorinRDS14}. 
In this section, we recall \acp{tgg}, the languages they define, and the operationalization of \ac{tgg} rules, which is the basis for using \acp{tgg} for advanced model management tasks like model translation or synchronization. 
We introduce these notions formally here, where we need the corresponding notation in \cref{sec:solutions}. 
Other notions are presented intuitively, and we provide references to the relevant literature for their formal definitions.
For readers unfamiliar with \acp{tgg}, we use our running example in \shortorlong{the long version of our paper~\cite{KK26}}{\cref{app:examples-tggs}} to illustrate each introduced concept. 

\paragraph{Variants of \acp{tgg} and their languages}
We define triple graphs using the formalism suggested by Golas et al.~\cite{GolasLEG12} that slightly deviates from the original definition from Schürr~\cite{Schurr94}; however, we use a set theory-based definition (instead a morphism-based one as Golas et al. do). 

A \emph{triple graph} is a \emph{typed graph} (i.e., it is possible to assign recurring \emph{roles} to its elements) consisting of three distinguished parts being subgraphs of the triple graph: the \emph{source}, the \emph{target}, and the \emph{correspondence graph}. 
Intuitively, the source graph captures a model in one modeling language (e.g., some design model of a system), the target graph captures a model in another modeling language (e.g., some abstracted view of the design model), and the correspondence graph serves as a structure to allow for tracing between both. 

\begin{definition}[Triple graph]
    A \emph{triple graph} is a tuple $\TrGraph{G}$, where $G$ is a graph and $\SrcGraph{G}, \CorrGraph{G}, \TarGraph{G}$ are subgraphs of $G$, referred to as \emph{source}, \emph{correspondence}, and \emph{target part}, such that:
    \begin{enumerate}
        \item The set of nodes $V_G$ of $G$ is a disjoint union of the sets of nodes $V_{\SrcGraph{G}}, V_{\CorrGraph{G}}, V_{\TarGraph{G}}$. 
        \item Every edge $e \in E_G$ of $G$ maximally belongs to one of the edge sets $E_{\SrcGraph{G}}$ or $E_{\TarGraph{G}}$; in particular, the edge set $E_{\CorrGraph{G}}$ is empty.
        \item Every edge $e \in E_G$ that neither belongs to $E_{\SrcGraph{G}}$ nor to $E_{\TarGraph{G}}$ is directed from a node from $V_{\CorrGraph{G}}$ to either a node from $V_{\SrcGraph{G}}$ or from $ V_{\TarGraph{G}}$.
    \end{enumerate}
\end{definition}
To further enhance the expressiveness of the considered models, triple graphs can be \emph{attributed}, i.e., nodes can carry values for various attributes. 
A general formalization of the attribution of graphs can be found in~\cite{EhrigEPT06}. 

\emph{\Aclp{tgg}} are a declarative, rule-based formalism that describe how two models from the source and the target domain co-evolve. 
This defines a correspondence-relation between pairs of models from the two domains by stipulating that two models are in correspondence when a triple graph can be derived, starting at the empty graph and using the rules of the given grammar, such that the one model is its source part and the other its target part. 
While the grammar defines correspondence on the level of models, the correspondence links in a triple graph define correspondence on the element level. 
\Ac{tgg} rules are \emph{monotone}, i.e., they only create structure and do not delete. 
Intuitively, a \ac{tgg} rule requires a certain pattern to exists, extends that pattern, but only does so if further specified forbidden pattern are not present. 
As advanced concepts, \ac{tgg} rules can manipulate attribute values of graph nodes~\cite{AnjorinVS12} and be equipped with so-called \emph{\acp{nac}} that restrict the context in which a rule is allowed to be applied~\cite{AnjorinST12}.

\begin{definition}[\Ac{tgg} rule. Application] 
    A \emph{(\ac{tgg}) rule} $\RuleFull$ consists of a \emph{plain rule} $r$, a set $\mathit{NAC}$ of \aclp{nac}, and a set $\mathcal{C}$ of \emph{attribute constraints}. 
    The plain rule $r$ is an injective morphism $\RuleMorphism$ between triple graphs $L$ -- the \emph{\ac{lhs}} -- and $R$ -- the \emph{\ac{rhs}} -- of the rule; the \ac{lhs} $L$ of a rule is also referred to as \emph{context} of the rule. 
    The set $\mathit{NAC}$ is a set of triple graphs such that $L$ is a subgraph of each of them. 
    The set $\mathcal{C}$ consists of Boolean expressions, regulating the attribute values appearing in the \ac{lhs} $L$ of the rule. 
    A rule $\rho$ is \emph{applicable} to a triple graph $G$ if there exists an injective morphism $m\colon L \hookrightarrow G$ that (i) cannot be extended to a morphism $q\colon N \hookrightarrow G$ for any of the graphs $N \in \mathit{NAC}$ and (ii) induces an evaluation on the attribute values of the \ac{lhs} $L$ such that every condition $c \in \mathcal{C}$ is satisfied by it. 
    \emph{Application} of a rule then means to extend $G$ by $R \setminus L$ (the node- and edge-wise difference of $L$ and $R$) at the location indicated by $m$, obtaining a triple graph $H$ as result; formally, this amounts to computing $H$ as a pushout of $m$ and $r$.
    Such an application, or \emph{transformation step} is denoted via $\TrafoStep{G}{H}{\rho}{m}$; a sequence of transformations via $\TrafoSeq{G}{H}$ or via $\TrafoSeqRuleSet{G}{H}{\mathcal{R}}$ if one wants to indicate that all transformation steps have been performed applying rules from a set $\mathcal{R}$.
\end{definition}

A further feature that has been developed for \acp{tgg} is \emph{multi-amal\-ga\-ma\-tion}. 
Being established in the area of graph transformation~\cite{GolasEH10}, Leblebici et al. transferred it to \acp{tgg}~\cite{LeblebiciAST17}. 
Multi-amal\-ga\-ma\-tion equips certain rule elements with a ``for-each''-like syntax, enabling one to concisely express that a certain operation should be performed as often as possible. 
A \emph{kernel rule} specifies elements that are to be matched or created (depending on which side of the rule they appear in) once, i.e., it is an ordinary \ac{tgg} rule. 
Additional \emph{multi-rules} extend this kernel rule by further context elements and elements to be created. 
Their application semantics is as follows: 
Given a match for the kernel rule, the kernel rule is applied exactly once at this match. 
The larger context of each multi-rule is matched as often as possible, extending the match of the kernel rule, and, for each such match, the elements declared to be created by the multi-rule are created at that match. 

\Acp{tgg} consist of sets of \ac{tgg} rules and are used to define various languages. 
Two models from the source and the target modeling languages are considered to be consistent to each other when there exists a triple graph in the language of the given \ac{tgg} such that the source and the target model are its source and target parts.
\begin{definition}[\Acl{tgg}. Language]
    A \emph{\acf{tgg}} consists of a set of \ac{tgg} rules $\mathcal{P}$. 
    The \emph{language} of a \ac{tgg} is the set of triple graphs derivable via the given rules from the empty graph, i.e., 
        $\LangTGG{\mathcal{P}} \coloneqq \{\TrGraph{H} \mid \TrafoSeqRuleSet{\emptyset}{H}{\mathcal{P}}\}$.
\end{definition}
Importantly, all features of \acp{tgg} we considered (\acp{nac}, attribute constraints, and multi-amalgamation) increase the expressiveness of \acp{tgg}, i.e., allow one to define languages that cannot be defined using just plain rules~\cite{WeidmannOR19}.

\paragraph{Operationalization of \ac{tgg} rules}
\Acp{tgg} as introduced so far describe how consistent models co-evolve; i.e., in our application scenario, they describe how to consistently co-evovle system models and views. 
The main usage scenario for \acp{tgg}, however, is not the \emph{construction} of correlated models but more challenging model management tasks like the \emph{translation} of a given model into one of the second domain or \emph{consistency restoration} after one or both previously corresponding models have been edited. 
For these tasks, \ac{tgg} rules can be (automatically) \emph{operationalized}.
The crucial operationalized rule variant for this paper are so-called \emph{forward rules}. 
A forward rule of a given \ac{tgg} rule adapts that rule in the following way: 
The context elements of the original rule remain context elements of the forward rule. 
Elements that the original rule creates on the correspondence or the target side also remain to be created. 
Elements that the original rule creates on the source side, however, become context. 
Furthermore, to keep track of already translated elements, these formerly created elements get marked. 
This marking process can be formalized in various different ways, e.g., via translation attributes~\cite{HermannEGO10}, via subrules~\cite{LeblebiciAFVS17}, or via dedicated sub-triple graphs~\cite{Kosiol22}. 
By and large symmetrically to the case of forward rules (except for the assignment of attribute values), \emph{backward rules} can be derived from \ac{tgg} rules. 
Instead of parsing structure on the source part of a triple graph and transferring it to the target part (as forward rules do), backward rules parse the target part and transfer the according structure to the source part.
In our application case in this paper, the forward rules are of crucial importance: Given a model of a system, they can be used to compute a view of it. 
In the future, we also want to use forward as well as backward rules to propagate changes between models and views. 

\subsection{Operator-Based View Definition}
\label{sec:background:operators}

Here, we give a brief overview of the model query operators defined by \citeauthor{konig_transforming_2025}~\cite{konig_transforming_2025}.
There, a view definition describes an asymmetric transformation between one or multiple source models and a projective view on these models.
At the same time, the view definition describes a transformation to create the view type, i.e., the metamodel of the resulting views.
A view definition consists of a set of model queries, which describe parts of the transformation.
For that, a model query projects a fixed set of related elements of the source models to a fixed set of elements in the view.

\begin{table}
    \caption{Overview of the basic query operators defined by \citeauthor{konig_transforming_2025}}
    \label{tab:operators}
    
    \centering
    \renewcommand{\arraystretch}{1.2}
    \begin{tabularx}{\columnwidth}{Xp{2cm}p{3.3cm}}
        \toprule
        \textbf{Name} & \textbf{Symbol} & \textbf{Arguments} \\
        \midrule
        Selection & $\sigma(P \Rightarrow Q)$ & $P$ Source pattern \newline $Q$ Target pattern \\ \addlinespace
        Filter & $\phi(p)$ & $p$ Filter predicate \\ \addlinespace
        Attribute\newline Projection & $\pi(x \Rightarrow y)$ \newline $\pi(y := f(\dots))$ & $x$ Source attribute \newline $y$ Target attribute \newline $f$ Projection function \\ \addlinespace
        Reference\newline Projection & $\rho(R \Rightarrow S)$ \newline $\rho_c(R \Rightarrow S)$ & $R$ Source reference pattern \newline $S$ Target reference pattern \\
        \bottomrule
    \end{tabularx}
\end{table}

To write view definitions, \citeauthor{konig_transforming_2025} introduced an operator-based notation for model queries.
We present an overview of the available operators in \cref{tab:operators}.
A model query starts with a single selection operator, which projects the source elements to the view, and can contain an arbitrary number of other operators in addition, a center dot~$\cdot$ indicating concatenation. 
For that, the selection operator $\sigma$ takes a source pattern and a target pattern.
The source pattern describes the model elements and their relations in the source models, while the target pattern describes the resulting elements in the view.
Elements of the models and the view are specified using the name of their metaclass.
\citeauthor{konig_transforming_2025} restricted source patterns to directed paths, but we accept any well-typed graph. 

In \cref{query:example} we give a small example of a model query based on the two metamodels in \cref{fig:example-simulink,fig:example-hardware}.
The relations in the source pattern can be explicit references, written as $\rightarrow_{\mathit{foo}}$ where $\mathit{foo}$ is the name of the reference, or implicit in form of a join, written as $\bowtie_{\mathit{bar}}$ where $\mathit{bar}$ is a boolean condition.
In the example in \cref{query:example}, we join a \modelElement{Block} with a \modelElement{Sensor} and, from those, create a \modelElement{Block} and a referenced \modelElement{DebugInfo} in the view (instead of a join condition we only write the shared attribute as a shorthand).
Joins between model elements behave similar to joins from relational algebra~\cite{codd_relational_1970}, but we will discuss some limitations in more detail in \cref{sec:problems:joins}.
\begin{align}
    \begin{split}
        &\sigma(\text{Block} \bowtie_{\text{sensorGroup}} \text{Sensor} \Rightarrow \text{Block} \rightarrow_{\text{debug}} \text{DebugInfo})\:\cdot\\
        &\phi(\text{Block.resolution} > 16)\:\cdot\\
        &\pi(\text{Sensor.resolution} \Rightarrow \text{DebugInfo.sensorRes})\:\cdot\\
        &\rho(\text{Block} \rightarrow_{\text{inPorts}} \text{InPort} \Rightarrow \text{Block} \rightarrow_{\text{in}} \text{Input})
    \end{split}
    \label{query:example}
\end{align}

Additionally, there is the filter operator $\phi$, which selects model elements based on a filter condition.
The transformation described by the model query is only executed if the filter condition evaluates to true on the model elements described by the source pattern of the selection.
In the example in \cref{query:example}, we only project blocks associated with sensor with a resolution greater $16$.

The view elements can be further populated by projecting attributes and references from the source models.
Attributes are projected using the attribute projection operator $\pi$, which either selects an attribute from the source elements or computes a new attribute.
The first mode is written as $A.a \Rightarrow B.b$ where $a$ is an attribute of the source element $A$ and $b$ is the resulting attribute in the target element $B$.
We do this in the example in \cref{query:example} to project the sensor attribute \emph{resolution} to the \emph{DebugInfo} element as attribute \emph{sensorRes}.
The second mode is written as $B.b := f(\dots)$ where $f$ is a function on the attributes of the source elements.

Similar, references are projected using the reference projection operator $\rho$.
While the source pattern of the selection operator can contain only 1-to-1 references to other model elements, the reference projection operator projects 1-to-$n$ references to model elements, which are transformed separately.
\citeauthor{konig_transforming_2025} limited this to only one 1-to-$n$ reference per source reference pattern. 
Instead, we define that the reference projection operator selects instances of the source reference pattern in the models. 
For each unique instance of the source reference pattern, an instance of the target reference pattern is created in the view. 
The source reference pattern may include elements from the source pattern of the query, while the target reference pattern creates exactly one reference from one of the elements in the target pattern.
In the example in \cref{query:example}, we project the reference \emph{inPorts} to the view, specify the view metaclass \emph{Input} as its target, and change its name to \emph{in}.
A special case of the reference projection operator is containment projection, written as $\rho_c$.
In this case, the referenced elements are only transformed if the containing element is transformed.

\section{Complex View Definition Operators} 
\label{sec:problems}

Beyond the basic operators recalled in \cref{sec:background:operators}, \citeauthor{konig_transforming_2025}~\cite{konig_transforming_2025} encountered a number of limitations of their model query operators.
Most of them exist because the necessary \ac{tgg} would be complicated to construct. 
In this section, we examine three of them, namely \emph{aggregation}, \emph{cross-product joins}, and \emph{overlapping queries}; in \cref{sec:solutions}, we propose solutions using advanced \ac{tgg} extensions.

\subsection{Aggregation}
\label{sec:problems:aggregation}

\begin{figure}
    \centering
    \includegraphics{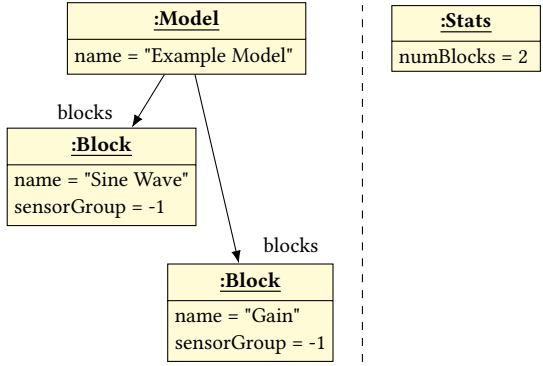}
    \caption{Part of a model (left) and a view (right) showing the aggregation over multiple referenced model elements.}
    \label{fig:view-aggregation}
\end{figure}

With the operators presented in \cref{sec:background:operators}, we are able to project existing attributes, as well as calculate new attributes from the values of the attributes of the model elements in the source pattern of a query.
However, as the source pattern has a constant size, we cannot calculate the value of an attribute from a set of model elements of unknown size, e.g., all elements referenced by an element in the source pattern.
What we would like to have is an aggregation function, similar to aggregation in extensions of relational algebra~\cite{codd_relational_1970}.
For that, we introduce the aggregation operator $\alpha$, which takes the application of an aggregation function as its first argument and the definition of a view attribute as its second argument. 
For the aggregation function we allow functions that incrementally build up the resulting value, i.e., functions that can be written as a fold over the aggregated elements.
This is similar to how aggregation is handled in relational query languages. 
We show an example of this in \cref{query:aggregation} using $count$ as an aggregation function to count the number of blocks in a model.
In \cref{fig:view-aggregation}, we show an example model instance and the view resulting from applying the query in \cref{query:aggregation} to it.
As the model references two blocks, the attribute \emph{numBlocks} of the view element \emph{Stats} is set to $2$.
\begin{align}
    \begin{split}
        &\sigma(\text{Model} \Rightarrow \text{Stats})\:\cdot\\ &\alpha(count(\text{Model} \rightarrow_{\text{blocks}} \text{Block}) \Rightarrow \text{Stats.numBlocks})
    \end{split}
    \label{query:aggregation}
\end{align}

Generating a \ac{tgg} for the aggregation is, however, difficult with traditional \ac{tgg} formalisms.
It is well-known that \acp{tgg} struggle to relate a structural encoding of information to an attribute-based one~\cite[Sect.~3.2]{AnjorinB26}.
We therefore consider this an interesting problem.

\subsection{Cross-Product Joins}
\label{sec:problems:joins}

\begin{figure}
    \centering
    \includegraphics{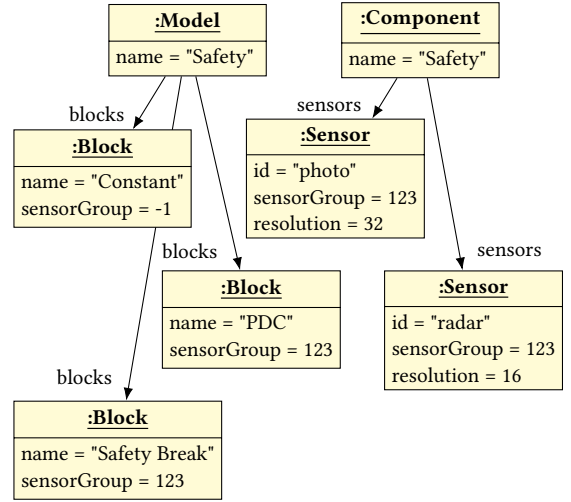}
    \caption{Part of two models (top) and a view (bottom) showing a conditional cross-product join of elements from the two models.}
    \label{fig:view-join}
\end{figure}

The selection operator $\sigma$, presented in \cref{sec:background:operators}, can be used to select elements related by explicit references or implicitly by joining two elements.
For the latter, this is done using a join condition, e.g., $\bowtie_{\text{Model.name} \: = \: \text{Component.name}}$.
However, \citeauthor{konig_transforming_2025}~\cite{konig_transforming_2025} restrict the join conditions to 1-on-1 joins, i.e., a model element can only participate once in the same join.
In general, of course, we want to be able to do full cross-product joins as well.
In our example in \cref{query:join}, we want to join blocks with their associated sensors to present developers with a list of blocks and sensors to check, as motivated in \cref{sec:example}.
For that, we do not want to transform the elements individually, but instead produce a \emph{Block} element as well as a \emph{Sensor} element for each associated pair in the source models.
We further restrict the selected pairs with a filter condition.
In \cref{fig:view-join}, we show the result of this join on a small example model.
Note that for a more complex model, the same \emph{Block} or \emph{Sensor} element could occur multiple times, in different joined pairs, in the view.
\begin{align}
    \begin{split}
        &\sigma(\text{Block} \bowtie_{\text{sensorGroup}} \text{Sensor} \Rightarrow\\
        &\hspace{2em}\text{Block} \rightarrow_{\text{source}} \text{Sensor})\:\cdot\\
        &\phi(\text{Sensor.resolution} > 16)
    \end{split}
    \label{query:join}
\end{align}

\subsection{Overlapping Queries}
\label{sec:problems:overlap}

When generating a \ac{tgg} from relational queries, \citeauthor{konig_transforming_2025}~\cite{konig_transforming_2025} make the assumption that queries in a view definition do not overlap, i.e., that they do not reference the same source model elements.
However, there certainly are cases where we would like to have overlapping queries.
For example, we could show the same element with different representations, as done in projectional editing, use model elements to highlight constraint violations, or present summary information.
We show queries \cref{query:overlap-1,query:overlap-3} as an example.
Note that we assume that \emph{InPort}, \emph{OutPort}, and \emph{DataType} elements are transformed separately with identical metaclass names and all attributes to simplify the example.
\begin{align}
    \begin{split}
        &\sigma(\text{Block} \Rightarrow \text{InternalBlock})\:\cdot\\
        &\phi(startsWith(\text{Block.name}, \text{``\_''}))
    \end{split}\label{query:overlap-1}\\[1.5em]
    \begin{split}
        &\sigma(\text{Block} \Rightarrow \text{SensorBlock})\:\cdot\\
        &\phi(\text{Block.sensorGroup} > -1)\:\cdot\\
        &\rho(\text{Block} \rightarrow_{\text{outPorts}} \text{OutPort} \Rightarrow \text{SensorBlock} \rightarrow_{\text{out}} \text{OutPort})
    \end{split}\label{query:overlap-2}\\[1.5em]
    \begin{split}
        &\sigma(\text{Model} \Rightarrow \text{ModelDiagnostics})\:\cdot\\
        &\rho(\text{Model} \rightarrow_{\text{blocks}} \text{Block} \rightarrow_{\text{inPorts}} \text{InPort} \rightarrow_{\text{type}} \text{DataType}\\
        &\hspace{2em}\Rightarrow\\
        &\hspace{2em}\text{ModelDiagnostics} \rightarrow_{\text{types}} \text{DataType})\:\cdot\\
        &\rho(\text{Model} \rightarrow_{\text{blocks}} \text{Block} \rightarrow_{\text{outPorts}} \text{OutPort} \rightarrow_{\text{type}} \text{DataType}\\
        &\hspace{2em}\Rightarrow\\
        &\hspace{2em}\text{ModelDiagnostics} \rightarrow_{\text{types}} \text{DataType})
    \end{split}\label{query:overlap-3}
\end{align}

\Cref{query:overlap-1,query:overlap-2} both select \emph{Block} elements, however, \cref{query:overlap-1} filters the result for internal blocks (which have names starting with an underscore), while \cref{query:overlap-2} filters for blocks that have sensors associated to them.
As there might be internal blocks that have sensors associated to them, a block from the source model might be transformed by none, either, or both of the queries.

\Cref{query:overlap-3} also selects \emph{Block} elements, but in a source reference pattern.
As the \emph{inPorts} reference in the source reference pattern is used here exclusively, we could create a reference \emph{types} in the view whenever a reference \emph{inPorts} is found in the source model.
Note that the \emph{Block} element, from which the \emph{inPorts} references originates, might or might not be transformed by another query.

Finally, \cref{query:overlap-3} projects the \emph{outPorts} reference to the \emph{types} reference in the view type.
As \cref{query:overlap-2} also projects the \emph{outPorts} reference, all parts of the source reference pattern in \cref{query:overlap-3} might or might not be transformed by other queries.

Overlapping queries are difficult to handle when generating \acp{tgg} for the transformation.
In common variants of \acp{tgg}, nodes are either context nodes or create nodes.
By that, whenever a node in a rule might or might not be created by another rule, we would need to duplicate the rule with a context node in one and a create node in the other rule.
As this can be the case for multiple nodes in a rule, we would require $2^n$ rules where $n$ is the number of nodes with unique overlaps to other queries.

\begin{figure}
    \centering
    \includegraphics{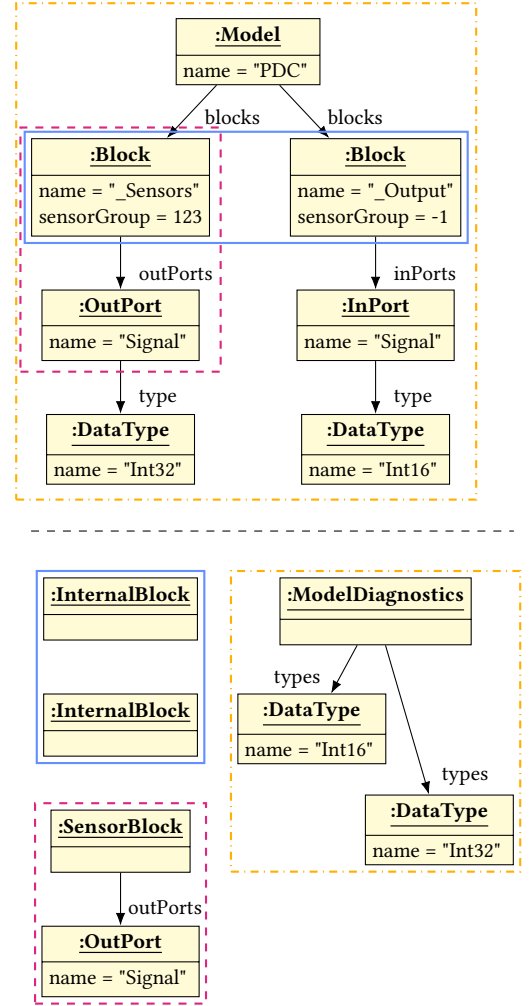}
    \caption{Part of a model (top) and a view (bottom) showing the result of the three overlapping Queries (\ref{query:overlap-1}, blue, solid), (\ref{query:overlap-2}, red, dashed), and (\ref{query:overlap-3}, yellow, dashed and dotted).}
    \label{fig:view-overlap}
\end{figure}

\section{Using and Extending Triple Graph Grammars}
\label{sec:solutions}
In this section, we sketch how we solve the problems introduced in the last section.

\subsection{Aggregation}
\label{sec:solutions:aggregation}
Support for aggregation has to be developed separately for each kind of aggregation function. 
For many basic aggregation functions, defining \ac{tgg} rules that co-evolve models consistently capturing the aggregation function is possible, just using means that are well-established for attributed graphs, namely using rules to set attribute values according to operations supported by an underlying algebra~\cite{EhrigEPT06}. 
The resulting rules, however, are such that they cannot always be operationalized in both directions, caused by the fact that regularly, the operations used to implement the aggregation are not invertible in a unique way. 
That is, we are able to operationalize such a \ac{tgg} rule into a forward rule but not (uniquely) into a backward rule. 
For our application scenario, this does not cause a too serious problem, because we are mainly interested in constructing views (and not in adapting a model according to an edited view).

Given an aggregation like the one displayed in \cref{query:aggregation}, our general approach to construct \ac{tgg} rules from it is as follows.
\begin{enumerate}
    \item We translate the selection operation $\sigma$ at the start of the query as already established in~\cite{konig_transforming_2025}. 
        Additionally, we scan the second argument of the aggregation operator $\alpha$ for the view attribute defined there and create and initialize this attribute at the respective target node.
        The default value for initialization depends on the aggregation function. 
        Additionally, depending on the aggregation function, the rule might create (but not yet initialize) a \emph{helper attribute} at the correspondence node it creates; this shall serve to facilitate the computation of the value of the view attribute later on.%
        \footnote{We position this helper attribute at the correspondence node because, usually, users do not interact with these and so we minimize diversion of users. It would also simply be possible to maintain these helper attributes as a separate data structure, but we opted for this more integrated approach.}

    \item We translate the aggregation operation into a second \ac{tgg} rule. 
        The \ac{lhs}, i.e., context of this rule is always the \ac{rhs} of the rule constructed under 1., except that we omit the assignment of attribute values. 
        Its \ac{rhs} is specific to the selection operation $\sigma$ and, in particular, the aggregation function to be translated. 
        On the source part, the rule always creates the structure, or assigns the attribute value, that is declared in the source pattern $P$ of the selection operation $\sigma$. 
        On the target part, the rule always updates the view attribute according to the aggregation function.
        Additionally, depending on the aggregation function, the rule might need to update the helper attribute of the correspondence node of its \ac{lhs}.
\end{enumerate}

\begin{figure}
    \centering
    \includegraphics{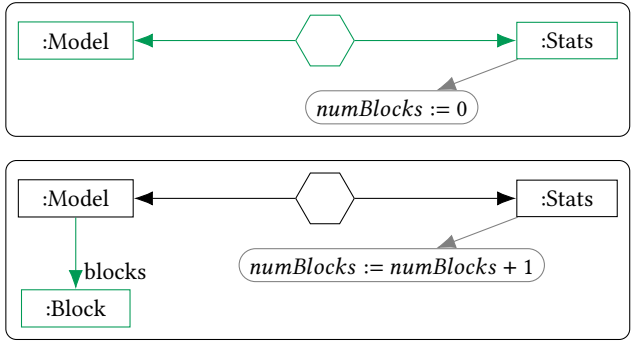}
    \caption{\ac{tgg} rules for the aggregation defined in \cref{query:aggregation}.}
    \label{fig:tgg-aggregation}
\end{figure}

\begin{example}[\Ac{tgg} rules for the \emph{count} function]
    \Cref{fig:tgg-aggregation} shows the two rules that arise for the example from \cref{query:aggregation}, i.e., for the \emph{count} function. 
    The first rule creates the correspondence between the source and the target pattern defined in the selection operator of the query, creates the view attribute $\mathit{numBlocks}$ and assigns to it the default value $0$ (as suitable for the \emph{count} function). 
    The second rule creates a \modelElement{Block} on the source part and increments the $\mathit{numBlocks}$ attribute by $1$.
\end{example}
In \shortorlong{the long version of our paper~\cite{KK26}}{\cref{app:aggregation}} we briefly sketch how we treat further common aggregation functions beyond the \emph{count} function. 

On the source part of a triple graph, the rules we derive for aggregation only match or create structure. 
We can therefore derive forward rules as usual, matching or marking as translated those elements on the source part instead of matching or creating them. 
On the correspondence and the target part, the elements are matched or created and attribute values assigned just as in the underlying \ac{tgg} rule. 
Backward rules often cannot be derived, because, generally, aggregation functions are not (uniquely) invertible. 
While this reinforces the observation made in~\cite[Section~3.2]{AnjorinB26} that \acp{tgg} are not well-equipped to synchronize in cases where information is represented as structure in one model but as attribute values in another, we do not see that this poses a problem to our application here, where we are foremost interested in just deriving the view, i.e., the more abstract model that might summarize structural information and represent it as a single numerical value, for instance.

\begin{figure}
    \centering
    \includegraphics{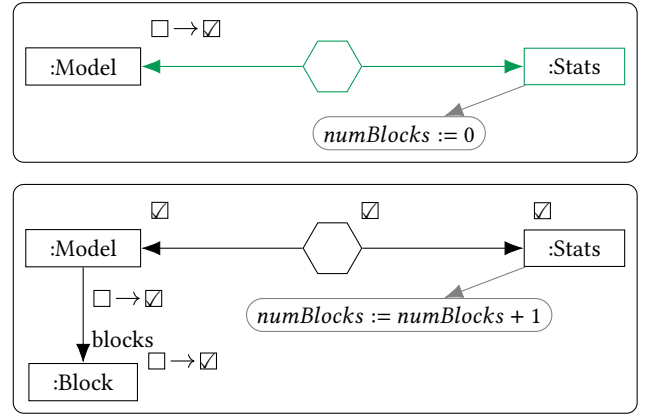}
    \caption{Forward rules of the \ac{tgg} rules in \cref{fig:tgg-aggregation}.}
    \label{fig:operation-aggregation}
\end{figure}

\begin{example}
    When forward operationalizing the \ac{tgg} rules for the count function displayed in \cref{fig:tgg-aggregation}, we obtain the forward rules displayed in \cref{fig:operation-aggregation}.
    The first rule marks an existing node of type \modelElement{Model} as translated on the source part and creates a corresponding node of type \modelElement{Stats} on the target part; moreover, it assigns the \emph{numBlocks} attribute to 0. 
    The second rule matches an already translated \modelElement{Model} node and its corresponding \modelElement{Stats} node on the target part, marks one of its previously untranslated \modelElement{Blocks} as now translated and increments the corresponding \emph{numBlocks} counter by 1.
\end{example}

\subsection{Cross-Product Joins}
\label{sec:solutions:joins}

\Acp{tgg} have been designed to be able to express $m \times n$-relations between elements on the source and the target part~\cite{Schurr94}. 
Here, however, as introduced in \cref{sec:problems:joins} we are interested in a more general problem, namely in computing a \emph{join} (i.e., an $m \times n$-relation between elements) on the source side and relating this join to a structure on the target part. 
To express the desired behavior compactly, we make use of multi-amalgamated \ac{tgg} rules. 
Given a selection query $\sigma(P \Rightarrow Q)$, where the source pattern $P = P_1 \bowtie_{c} P_2$ is defined as a join with join condition $c$, we derive two rules.

The first rule has as kernel rule a rule that just creates the source pattern $P_1$. 
Depending on the join condition $c$, the kernel rule might also create attributes for nodes of $P_1$ and assign these to variables. 
This kernel rule is extended by a single multi-rule, which matches the source pattern $P_2$ and creates the target pattern $Q$ and a correspondence node that connects to each node of $P_1, P_2$ and $Q$. 
Moreover, the multi-rule is equipped with an attribute condition (having access to the variables used in the kernel rule) that ensures that it is only applicable when the join condition $c$ is met.
The second rule is completely analogous just that the roles of $P_1$ and $P_2$ are switched.

\begin{figure}
    \centering
    \includegraphics{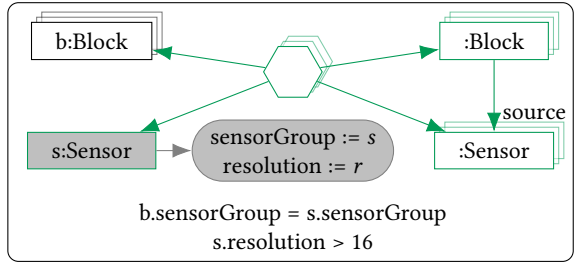}
    \caption{%
        \ac{tgg} rule for the left side of an inner join, as defined in \cref{query:join}.
    }
    \label{fig:tgg-inner-join}
\end{figure}

\begin{example}
    \Cref{fig:tgg-inner-join} shows the rule we derive for the left side of the join from \cref{query:join}. 
    This rule ensures the completeness of the considered join by making use of multi-amalgamation: 
    A newly created \modelElement{Sensor} is joined with each already existing \modelElement{Block} (of the same sensor group) and together these are put into correspondence with a pair of \modelElement{Block} and \modelElement{Sensor} on the target side.
    We show the rule for the right side of the join in \shortorlong{the long version of our paper~\cite{KK26}}{\cref{app:rule-operationalizations}}.
\end{example}

Operationalizing multi-rules to obtain forward and backward rules is treated in~\cite{LeblebiciAST17} and we can use these results without further adaptation. 
Intuitively, the forward rule we obtain from the rule in our example (\cref{fig:tgg-inner-join}) matches a yet untranslated \modelElement{Sensor}, translates it and joins that \modelElement{Sensor} to each already translated \modelElement{Block}, correlating this pair to a newly created pair of \modelElement{Sensor} and \modelElement{Block} on the target part. 
We depict this rule (and also the forward rule derived for the right side of the join) in \shortorlong{the long version of our paper~\cite{KK26}}{\cref{app:rule-operationalizations}}. 

\subsection{Overlapping Queries: An Extended Skip Semantics}
\label{sec:solutions:overlap}
Kosiol et al.~\cite{kosiol_finding_2023} have developed a skip semantics for graph transformation rules that we want to use to address the problem of overlapping queries. 
The main idea is that the actions of a rule are split into two classes: \emph{mandatory} ones that have to be performed upon application of the rule and \emph{potential} ones that are skipped if the outcome that the action would achieve is already present in the model. 
That is, for monotone rules as is the case for \acp{tgg}, a potential creation of an element is just skipped if a suitable element already exists in the model. 
This general idea can directly be transferred to the ordinary \ac{tgg} rules that make up the grammar. 
However, we have to newly develop a concept of how to operationalize this skip semantics and ensure the termination of the resulting operationalized rules. 
It turns out that this is easiest when further extending the original skip semantics with counters that keep track of how often an element has been skipped.

In the following, we develop this idea and explain how we use it to translate overlapping queries into \ac{tgg} rules. 
For this, we first introduce the concept of a \emph{subrule}. 
Intuitively, in our case, a subrule comes without an application condition, matches the same context elements, but potentially creates less elements than its encompassing rule.

\begin{definition}[Subrule of a \ac{tgg} rule]
    Given a \ac{tgg} rule $\RuleFull$, a \emph{subrule} of $\rho$ is a rule $\rho' = (r'\colon L' \hookrightarrow R')$ such that $L = L'$ and there is an injective morphism $i^{R'}_R \colon R' \hookrightarrow R$ such that $r = i^{R'}_R \circ r'$.
\end{definition}

The intuition behind rules with skip semantics is the following. 
First, a subrule captures the mandatory actions; hence, the subrule is applied as is. 
A rule that encompasses the subrule captures the additional potential actions (only creations in our case). 
Before applying this larger rule, it is first checked which of the to be created elements can be considered as already existent.
The according creation actions are then skipped and only the remaining ones performed. 
For this, we just reuse the definition of effect-oriented rules and transformations from~\cite{kosiol_finding_2023}. 
However, extending~\cite{kosiol_finding_2023}, NACs are allowed to not just extend the \ac{lhs} of the rule but also elements that are potentially to be created (because these might become context elements of the rule upon application). 
Moreover, the derivation of operationalized \ac{tgg} rules with skip semantics that we sketch at the end of this section is a new contribution. 
Formally, the skip semantics rests on a factorization of the morphism that embeds the subrule into the ambient rule.
\begin{definition}[\ac{tgg} rule with skip semantics]
    A \emph{\ac{tgg} rule with skip semantics} is a pair of rules $\rho_{\mathrm{skip}} = (\rho' = (r'\colon L \hookrightarrow R'), \RuleFull)$, i.e., $\rho'$ is a subrule of $\rho$, such that each graph $N \in \mathit{NAC}$ contains $L$, but maybe also nodes from $R \setminus R'$.
    Given a match $m\colon L \hookrightarrow G$ for the rule $\rho$, an \emph{application} of $\rho_{\mathrm{skip}}$ is defined as follows.
    \begin{enumerate}
        \item One searches for the largest (in terms of numbers of elements) factorization $i^{L}_{\Lind}\colon L \hookrightarrow \Lind$, $i^{\Lind}_R\colon \Lind \hookrightarrow R$ of $r$ such that (i) $\Lind$ does not contain any element from $R' \setminus L$ (i.e., no mandatory creations) and (ii) there exists an injective morphism $\mind\colon \Lind \hookrightarrow G$ with $\mind \circ i^{L}_{\Lind} = m$; this leads to an \emph{induced rule} $i^{\Lind}_R\colon \Lind \hookrightarrow R$ with \emph{induced match} $\mind\colon \Lind \hookrightarrow G$.
        \item Every negative application condition $N\in \mathit{NAC}$ is checked: 
        If $N$ contains a skip node from $R \setminus \Lind$, $N$ is just dropped. 
        Otherwise, one computes $N' \coloneqq N \cup \Lind$, and $N'$ becomes part of the negative application condition of the induced rule. 
        \item Finally, one applies the induced rule $i^{\Lind}_R\colon \Lind \hookrightarrow R$ at the induced match $\mind$, checking the newly computed negative application condition. 
        If that is violated, the overall application is stopped.
    \end{enumerate}
\end{definition}
Note that the application semantics is not deterministic as, in general, there might not exist a unique largest graph $\Lind$ for factorizing $r$. 
In our applications, so far, we did not observe any problems caused by that. 
We first describe the rules with skip semantics that we construct and later illustrate the concept using these rules.

Given a sequence of queries, we iteratively construct the following rules with skip semantics:
\begin{enumerate}
    \item For each query, we generate \ac{tgg} rules as developed in~\cite{konig_transforming_2025} and recalled in \cref{sec:background:operators} or introduced in \cref{sec:problems:aggregation,sec:problems:joins}. 
    This provides us with a complete set of operators defining consistency between model and view on the level of the individual queries; however, some queries might overlap.

    \item For each pair of rules $(\rho_1, \rho_2)$ we calculate the intersection
    \begin{equation*}
        R' \coloneqq (R_1 \setminus L_1) \cap (R_2 \setminus L_2)
    \end{equation*}
    and extend $\rho_1$ and $\rho_2$ to the \ac{tgg} rules with skip semantics
    \begin{equation*}
        \rho_i^{\mathrm{skip}} = (L_i \hookrightarrow (R_i \setminus R'), \rho_i) .\footnote{%
            If one or both of the rules participating in the intersection are multi-amalgamated rules, we separately compute those intersections for the kernel and the multi-rule, resulting in a multi-amalgamated rule where the kernel as well as the multi-rule are allowed to be equipped with skip semantics. 
            While we do not yet formalize this procedure in this paper, we are confident that this can be done.
        }
    \end{equation*}
    That is, the elements commonly created by both rules get equipped with the skip semantics. 

    \item To later ensure termination of the operationalized rules, we inspect the resulting rules with skip semantics to ensure that every such rule has a mandatory action on the source part, i.e., we search for rules $\rho^{\mathrm{skip}} = (\rho' = (r'\colon L \hookrightarrow R'), \rho= (r\colon L \hookrightarrow R, \mathcal{C}, \mathit{NAC}))$ where $L_{\mathrm{S}} = R'_{\mathrm{S}}$ (that is, on the source part, the \ac{lhs} of the subrule coincides with the \ac{rhs} of the second rule). 
    For any such rule, we use the correspondence node that $\rho$ creates between the source and target (reference) patterns of its underlying query, ensure that this is of a unique type, and add an additional NAC to $\rho$ that forbids that the elements from the underlying source (reference) pattern are already connected to a correspondence node of that type.
\end{enumerate}

\begin{figure}
    \centering
    \includegraphics{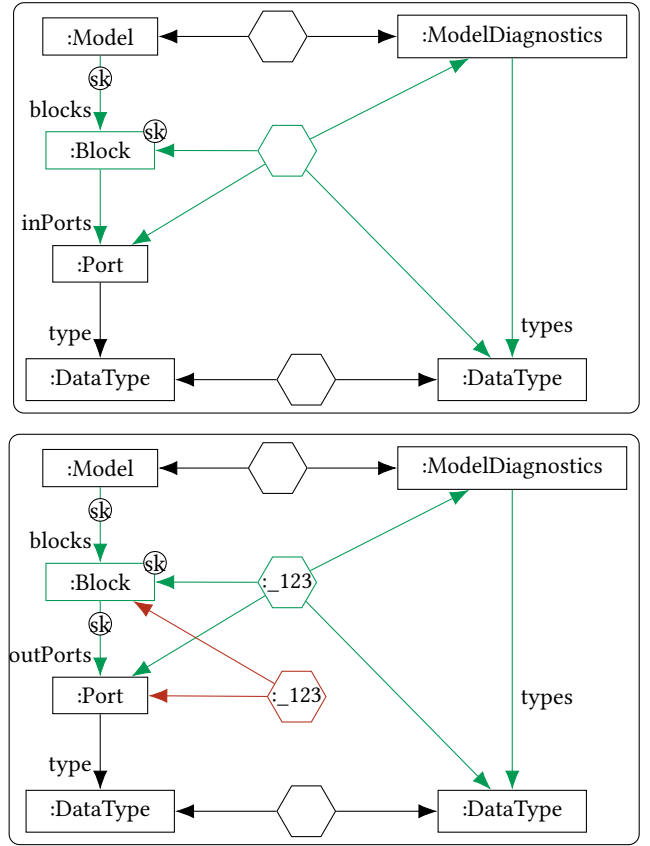}
    \caption{%
        \ac{tgg} rules for a query that overlaps with other queries, as defined in \cref{query:overlap-3}.
    }
    \label{fig:tgg-overlap}
\end{figure}

\begin{example}
    \Cref{fig:tgg-overlap} shows two of the rules we derive for \cref{query:overlap-1,query:overlap-2,query:overlap-3}; the elements marked by \enquote{sk} are the potential creations allowed to be skipped.  
    The two rules originally stem from the rules derived to translate the two reference projections from \cref{query:overlap-3}. 
    Because both rules create \modelElement{Blocks} and their incoming edges, these elements get equipped with the skip semantics.
    Moreover, the second rule also overlaps with the rule derived to translate the reference projection from \cref{query:overlap-2}; hence, also the \modelElement{outPorts}-edge receives the skip semantics. 
    In this way, the second rule ends up without any mandatory creation on its source part. 
    We therefore ensure that the correspondence node that the rule creates has a unique type and add a NAC that forbids the concerned source elements to already be connected to such a correspondence node. 
    Note that the referenced \modelElement{Block} is an element with skip semantics.
    Thus, the NAC just gets dropped whenever the \modelElement{Block} is actually created by an application of that rule, but gets evaluated whenever the \modelElement{Block} is skipped, i.e., matched to an already existing \modelElement{Block}. 
    In this way, we prevent the rule from being applicable multiple times to the same set of \modelElement{Model}, \modelElement{Block} and \modelElement{Port}.

    Regarding non-determinism, as soon as a \modelElement{Model} already has several \modelElement{Blocks}, it is not unambiguous which of these is used to compute the induced rule. 
    However, the resulting rules are isomorphic -- just their matches differ. 
    Importantly, each guarantees that a \modelElement{Block} exists after its application.
    In our use case in this paper (deriving views from models), this non-determinism does not pose any problems, because we apply the forward rules obtained from those rules as long as possible. 
    That is, the non-determinism affects the order in which \modelElement{Blocks} are translated to the view but not the overall result.
    In the future, we intend to formally clarify the general conditions under which this holds.
\end{example}

From rules with skip semantics, we derive the following forward rules: 
Context elements and mandatory creations are operationalized as usual, i.e., on the source part these become elements that already need to be translated and elements that get translated by the rule application, respectively. 
On the correspondence and target parts, elements remain context elements and mandatory creations, respectively.
Elements with skip semantics, in contrast, are treated as follows. 
On the correspondence and target parts, these elements keep their semantics, i.e., remain elements with skip semantics.
On the source part, however, skip elements either translate a previously untranslated element or match an element that already has been translated. 
In our example, this leads to forward rules that match already translated pairs of corresponding \modelElement{Model} and \modelElement{ModelDiagnostics} nodes and \modelElement{DataType} nodes, respectively, and, moreover, on the source part match a path from \modelElement{Model} to \modelElement{DataType}, across a \modelElement{Block} and a \modelElement{Port}, where this path might already be translated partially; in that, matched untranslated elements get translated. 
The rules then create a correspondence node that connects the matched \modelElement{Block} and \modelElement{Port} to the matched \modelElement{ModelDiagnostics} and \modelElement{DataType} nodes on the target part and introduces a \modelElement{types}-edge between the latter two. 
These rules are depicted in \shortorlong{the long version of our paper~\cite{KK26}}{\cref{app:rule-operationalizations}}.

\section{Conclusion}
\label{sec:conclusion}

In this paper, we continue the work of \citeauthor{konig_transforming_2025}~\cite{konig_transforming_2025} of providing a formal semantics to the NeoJoin view definition language by translating NeoJoin queries into \ac{tgg} rules. 
We address queries that had been omitted in~\cite{konig_transforming_2025} for their complexity. 
We find that -- by and large -- \acp{tgg} are well-suited for our task; however, we need to make use of advanced features of \acp{tgg} like negative application conditions, attribution and attribute conditions, and multi-amalgamation. 
As central technical contribution, to be able to deal with overlapping queries, we introduce a \emph{skip semantics} for \ac{tgg} rules, based on previous work by \citeauthor{kosiol_finding_2023}~\cite{kosiol_finding_2023}. 

In the future, there are two avenues to continue our work.
First, \acp{tgg} have regularly been used to obtain formal guarantees for model management processes (like model translation or synchronization) based on them~\cite{AnjorinLS15,HEOCDXGE15,FritscheKST21,Kosiol22}. 
Based on this, we intend to work out important properties of the views we derive. 
Given our experience with and existing research on \acp{tgg}, we are confident that we will be able to prove the following (or at least find criteria under which the following holds):
\begin{enumerate*}[(i)]
    \item The rules for aggregation behave correctly, i.e., aggregate attribute values or count elements as one would intuitively expect.
    \item The rules for joins behave correctly, i.e., construct the complete join. 
    \item The rules for overlapping queries behave correctly, i.e., each element is constructed exactly once. 
    \item The set of forward rules derived from all constructed rules is \emph{confluent}, i.e., using them to actually compute views can be done by applying the set of rules as long as possible in arbitrary order.
    \item User edits of a model or a view (except for manipulating aggregated attribute values) can be incrementally synchronized in either direction.
\end{enumerate*}
The mentioned criteria could include limiting the expressiveness of the described operators, such as restricting the source patterns of the selection operator.
While we do not yet prove results of this kind in this paper, developing a formal semantics as we do here makes such results tractable in the first place.

Second, and more practically, we intend to implement the concepts developed in this paper and want to extend the derivation of views, which we develop in this paper, by also synchronizing between models and views after one or both of them have been altered. 
Recent work on automatically generating repair operators for \acp{tgg}~\cite{FritscheKMST20,FritscheKST21,Kosiol22,FritscheKLMS24} provides a solid foundation to achieve this in an incremental and information-preserving fashion.
\newpage

\begin{acks}
This work builds on discussions from the NII Shonan Meeting No. 231 on Bidirectional Transformations: Foundations and Applications (bx). 
This work was supported by funding from the pilot program Core Informatics at KIT (KiKIT) of the Helmholtz Association (HGF) and funded by the Deutsche Forschungsgemeinschaft (DFG, German Research Foundation) --- SFB 1608 --- 501798263. 
We would like to thank Erik Burger for his helpful comments on an earlier version of this paper.
\end{acks}

\bibliographystyle{ACM-Reference-Format}
\bibliography{paper-detailed}

\shortorlong{}{%
    \clearpage
    \appendix
    \crefalias{section}{appendix}
    \section{Examples for Notions from Triple Graph Grammars}
\label{app:examples-tggs}

In this section, we use our running example to comprehensively illustrate \acp{tgg} and their established extensions as briefly introduced in \cref{sec:background:tgg}. 
We begin with an example for triple graphs. 
\begin{example}[Triple graphs]
    \Cref{fig:view-aggregation,fig:view-join,fig:view-overlap} can be interpreted as showing triple graphs in a simplified fashion. 
    The part above or left of the dashed line is always the source part of the triple graph, the part below or right of it the target part. 
    The source part is a system model defined over a combination of the two metamodels introduced in \cref{sec:example}, the target part is a view of that model. 
    We omit the correspondence parts because these would also be hidden from a user. 
    Formally, for example, in \cref{fig:view-aggregation} there also exists a correspondence node that has two outgoing edges leading to the \modelElement{Model} and the \modelElement{Stats} nodes, respectively. 
    Note that all examples make use of typing and attribution of graphs.
\end{example}

The next example illustrates \ac{tgg} rules.
\begin{example}[\Ac{tgg} rules]
    \Cref{fig:tgg-aggregation,fig:tgg-inner-join,fig:tgg-overlap} depict various \ac{tgg} rules, using a simple graphical syntax. 
    Black parts indicate the \ac{lhs}'s of rules, i.e., the context elements that have to be matched to apply a rule; an empty \ac{lhs} indicates that a rule is always applicable. 
    Green parts are the elements from $R \setminus L$, i.e., the elements that are to be created. 
    The rectangular elements on the left side of each rule always belong to the source parts of the graphs defining the rule, rectangular elements on the right to its target parts, and hexagon shaped elements are correspondence nodes. 
    Statements in rounded, gray boxes define variable assignments (formally, these belong to the \ac{rhs} of a rule and express that an attribution edge of the type indicated by the left part of the assignment is created that points to the attribute value indicated by the expression in the right part of the assignment).
    Thus, the second rule in \cref{fig:tgg-aggregation} declares to match a \modelElement{Model}- and a \modelElement{Stats}-node that are in correspondence and to create a \modelElement{Block}-node for the \modelElement{Model} and increment the \emph{numBlocks} counter by one. 
    The equations under the two rules from \cref{fig:tgg-inner-join} are attribute conditions, expressing that the rules are only applicable to \modelElement{Blocks} and \modelElement{Sensors} of the same sensor group and to \modelElement{Sensors} of high enough resolution.
    We later explain the gray fill and shadowed boxes appearing in the rules from this figure. 
    Finally, the red elements appearing in the second rule from \cref{fig:tgg-overlap} constitute a \emph{negative application condition} (NAC). 
    The NAC forbids this rule to be applied to a pair of \modelElement{Block} and \modelElement{Port} nodes that are already connected via an appropriately typed correspondence node.
    Again, we later explain the annotation of created elements via ``sk'' appearing in this figure.
\end{example}

The following example illustrates multi-amalgamated \ac{tgg} rules. 
\begin{example}[Multi-amalgamated \ac{tgg} rules]
    \Cref{fig:tgg-inner-join} displays two rules making use of multi-amalgamation. 
    The elements filled gray belong to the kernel rule, meaning that they are matched or created (depending on whether they belong to the \ac{lhs} or the \ac{rhs} of the rule) once. 
    Elements with shadows belong to the multi-rule, meaning that they are matched as often as possible (in case they belong to the \ac{lhs}) or created once per match of the multi-rule (in case they belong to the \ac{rhs}). 
    The attribute condition is evaluated for the multi-rule. 
    Summarizing, the first rule in \cref{fig:tgg-inner-join} creates one sensor on the source side, sets is sensor group and resolution (according to user provided or default values) and for each \modelElement{Block} existing on the source side and having a sensor group equal to the one of the created \modelElement{Sensor}, it creates one \modelElement{Block}, \modelElement{Sensor} and reference between them on the target side and connects all four nodes via a newly created correspondence node. 
\end{example}

Finally, we illustrate operationalized and, in particular, forward rules.
\begin{example}[Operationalized \ac{tgg} rules]
    \Cref{fig:operation-aggregation,fig:operation-inner-join,fig:operation-overlap} show the \emph{forward rules} that are derived from the \ac{tgg} rules displayed in \cref{fig:tgg-aggregation,fig:tgg-inner-join,fig:tgg-overlap}. 
    That an element is annotated with a check mark means that it was a context element of the underlying \ac{tgg} rule and now is only allowed to be matched to elements that already have been ``checked'', i.e., have already been translated by the previous application of a forward rule.
    (In that, elements that have been created by a forward rule on the correspondence or the target side are interpreted as ``checked'').
    If an element is annotated with an arrow transferring an empty box to a checked one, this means that this element is a source element that was to be created by the underlying \ac{tgg} rule. 
    Now, it is only allowed to be matched to source elements that have not yet been ``checked'' (translated) and checks such an element. 
    In this way, forward rules can be used to translate a source model to a target model (also building the correspondence structure in between).
    That is, in our application case, forward rules can be used to compute views from (families of) models. 
    The ``checking'' of elements that is necessary to keep track of which elements already have been translated can be formalized (and implemented) in various different ways.
\end{example}

\section{Examples for Additional and Operationalized Rules}
\label{app:rule-operationalizations}
In this section, we show rules that we explain but do not display in the main part of our paper. 
For the aggregation example, all constructed \ac{tgg} rules as well as their derived forward rules are displayed in the paper (\cref{fig:tgg-aggregation,fig:operation-aggregation}). 

\begin{figure}
    \centering
    \includegraphics{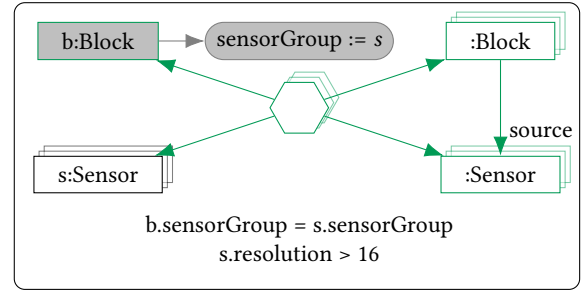}
    \caption{%
        \ac{tgg} rule for the right side of an inner join, as defined in \cref{query:join}.
    }
    \label{fig:tgg-inner-join-right}
\end{figure}

In \cref{fig:tgg-inner-join-right}, we show the \ac{tgg} rule for the right side of an inner join, in addition to the \ac{tgg} rule for the left side of an inner join, which we showed in \cref{fig:tgg-inner-join}.

\begin{figure}
    \centering
    \includegraphics{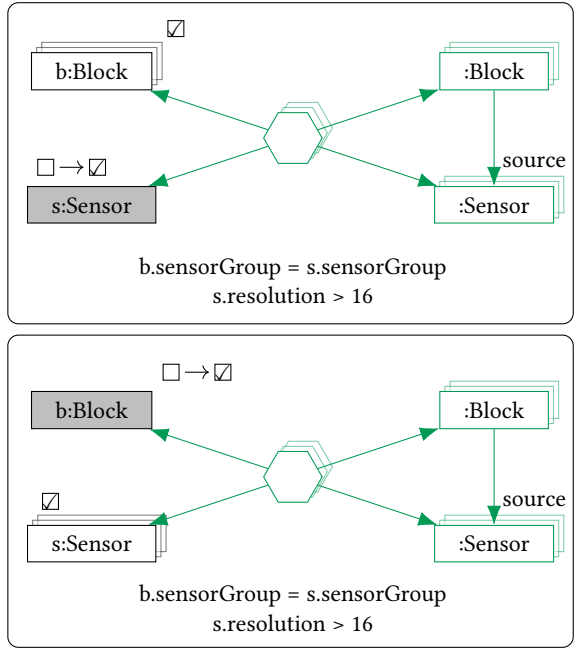}
    \caption{%
        Forward operationalization of the multi rules for an inner join, as defined in \cref{fig:tgg-inner-join}.
    }
    \label{fig:operation-inner-join}
\end{figure}

Furthermore, \cref{fig:operation-inner-join} shows both forward rules we derive from these two \ac{tgg} rules.
As briefly explained at the end of \cref{sec:solutions:joins}, these forward rules translate a \modelElement{Sensor} or \modelElement{Block}, respectively, by joining it with each previously translated \modelElement{Block} or \modelElement{Sensor}, respectively, by creating the required target pattern (once per pair of \modelElement{Block} and \modelElement{Sensor}) and connecting everything via a newly created correspondence node.

Finally, \cref{fig:operation-overlap} shows the two forward rules we derive from the \ac{tgg} rules with skip semantics we introduce for the case of overlapping queries (\cref{fig:tgg-overlap}); we briefly explained these rules at the end of \cref{sec:solutions:overlap}.
The notation \matchOrTranslate{} indicates that an element is either allowed to be matched to an already translated or to a yet untranslated element. 
This means, formally, these rules depict rule schemata and encode several rules. 
This reflects the fact that the underlying \ac{tgg} rules can, for example, either match or create a node of type \modelElement{Block}. 

\begin{figure}[b]
    \centering
    \includegraphics{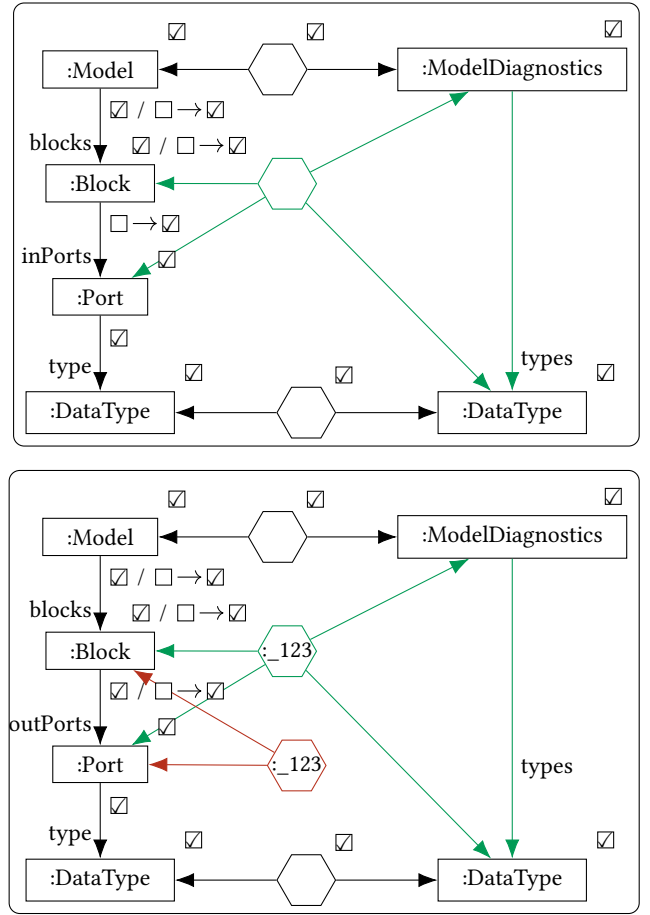}
    \caption{%
        Forward operationalization of the \acp{tgg} rules for overlapping queries with skip semantics, as defined in \cref{fig:tgg-overlap}.
    }
    \label{fig:operation-overlap}
\end{figure}

\section{Details on the Treatment of Aggregation}
\label{app:aggregation}

\Cref{tab:overview-translation-aggregation} provides an overview of how we treat a range of common aggregation functions. 
Note that, whereas \emph{Count} counts the number of graph elements of a certain type, the other aggregation functions read out some numerical attribute and perform a computation based on that. 
Depending on the kind of computation, we have to maintain a helper attribute that, in the worst case, keeps track of all attribute values encountered so far (e.g., in the cases of \emph{Median} and \emph{Mode}).  
The \emph{Count} function is the only one that is uniquely invertible: here, the view attribute contains all information necessary to reconstruct the source structure from it. 
From an average, however, one can neither derive from how many nor from which exact values it is computed. 
For the other aggregation functions this is typical. 
    
\begin{table*}
    \centering
    \caption{Overview of the translation of aggregation functions; $\readAttr$ refers to the attribute read out by the source pattern (if applicable),  $\helperAttr$ to the helper attribute, and $\viewAttr$ to the view attribute.}
    \label{tab:overview-translation-aggregation}
    \begin{tabular}{llccccc}
        \toprule
        \textbf{Function}   & \textbf{Type of} $\helperAttr$ & \textbf{Default} $\helperAttr$ & \textbf{Update} $\helperAttr$ & \textbf{Default} $\viewAttr$ & \textbf{Update} $\viewAttr$ & Invertible \\
        \midrule
         \emph{Average} & Integer & 0 & $\helperAttr \coloneqq \helperAttr + 1$ & 0 & $\viewAttr \coloneqq \frac{(\helperAttr-1) \times \viewAttr + \readAttr}{\helperAttr}$ & no  \\
         \emph{Count}       & ---               & --- & --- & 0 & $\viewAttr \coloneqq \viewAttr + 1$ & yes \\
         \emph{Max}     & ---               &  ---  & --- & $-\infty$ & $\viewAttr \coloneqq \max (\viewAttr, \readAttr)$ & no \\
         \emph{Median}      & Multiset          & $\emptyset$ & $\helperAttr \coloneqq \helperAttr \cup \{\readAttr\}$ & not set & $\viewAttr \coloneqq \Median(\helperAttr)$ & no \\
         \emph{Mode}        & Multiset          & $\emptyset$ & $\helperAttr \coloneqq \helperAttr \cup \{\readAttr\}$ & not set & $\viewAttr \coloneqq \Mode(\helperAttr)$ & no \\
         \emph{Range}       & Pair of numbers  & $(\infty, -\infty)$ & $\helperAttr = (\min(\helperAttr_0, \readAttr), \max(\helperAttr_1, \readAttr))$ & $-\infty$ & $\viewAttr \coloneqq \helperAttr_1 - \helperAttr_0$ & no \\
         \emph{Sum}         & ---               &  ---  & --- & $0$ & $\viewAttr \coloneqq \viewAttr + \readAttr$ & no \\
         \bottomrule
    \end{tabular}
\end{table*}

}

\end{document}